\documentclass[useAMS,usenatbib,usegraphicx]{mn2e}
\def\kms{km s$^{-1}$}

\def\aap{A\&A}
\def\apjl{ApJ}
\def\apj{ApJ}

\def\aj{AJ}
\def\mnras{MNRAS}

\def\pasp{PASP}

\title[A Gemini Snapshot Survey for DDs]
{A Gemini Snapshot Survey for Double Degenerates\thanks{Based on observations
obtained at the Gemini, MMT, and the Apache Point observatories. Gemini is operated by the
Association of Universities for Research in Astronomy, Inc., under a cooperative agreement
with the NSF on behalf of the Gemini partnership: the National Science Foundation
(United States), the National Research Council (Canada), CONICYT (Chile), the Australian
Research Council (Australia), Minist\'{e}rio da Ci\^{e}ncia, Tecnologia e Inova\c{c}\~{a}o
(Brazil) and Ministerio de Ciencia, Tecnolog\'{i}a e Innovaci\'{o}n Productiva (Argentina).
The MMT is a joint facility of the Smithsonian Institution and the University of Arizona.
The Apache Point Observatory 3.5-meter telescope is owned and operated by the Astrophysical
Research Consortium. This paper includes data taken at The McDonald Observatory of The
University of Texas at Austin.}}
\author[M. Kilic et al.]
       {Mukremin Kilic$^{1}$,
       Warren R. Brown$^2$,
       A. Gianninas$^1\thanks{Visiting astronomer, Kitt Peak National Observatory, National Optical Astronomy Observatory, which is operated by the Association of Universities for Research in Astronomy (AURA) under a cooperative agreement with the National Science Foundation.}$,
       Brandon Curd$^{1,3}$,
       \newauthor
       Keaton J. Bell$^{4}$,
       and Carlos Allende Prieto$^{5,6}$\\
       $^1$Department of Physics and Astronomy, University of Oklahoma, 440 W. Brooks St., Norman, OK, 73019, USA\\
       $^2$Smithsonian Astrophysical Observatory, 60 Garden St, Cambridge, MA 02138, USA\\
       $^3$Department of Astronomy, Harvard University, 60 Garden St, Cambridge, MA 02138, USA\\
       $^4$Department of Astronomy, University of Texas at Austin, Austin, TX 78712, USA\\
       $^5$Instituto de Astrof\'{\i}sica de Canarias, E-38205 La Laguna, Tenerife, Spain\\
       $^6$Departamento de Astrof\'{\i}sica, Universidad de La Laguna, E-38206 La Laguna, Tenerife, Spain
}

\begin{document}

\maketitle

\begin{abstract}

We present the results from a Gemini snapshot radial-velocity survey of 44 low-mass
white dwarf candidates selected from the Sloan Digital Sky Survey spectroscopy.
To find sub-hour orbital period binary systems, our time-series spectroscopy 
had cadences of 2 to 8 min over a period of 20-30 min. Through follow-up
observations at Gemini and the MMT, we identify four double degenerate binary systems
with periods ranging from 53 min to 7 h. The shortest period system, SDSS
J123549.88+154319.3, was recently identified as a subhour period detached binary by
Breedt and collaborators. Here we refine the orbital and physical parameters of this system.
High-speed and time domain survey photometry observations do not reveal eclipses or other
photometric effects in any of our targets. We compare the period distribution of these
four systems with the orbital period distribution of known double white dwarfs; the median
period decreases from 0.64 to 0.24 d for $M=0.3-0.5 M_{\odot}$ to $M<0.3 M_{\odot}$
white dwarfs. However, we do not find a statistically significant correlation between
the orbital period and white dwarf mass.

\end{abstract}


\section{Introduction}

There are now more than 90 short period binary white dwarfs known
\citep[e.g.,][]{saffer88,bragaglia90,marsh95,moran97,maxted00,morales05,nelemans05,vennes11,rebassa17,breedt17},
including more than three dozen systems that will merge within a Hubble time.
The majority of the merger systems were found in the last 7 years, thanks to the
Extremely Low Mass Survey \citep[the ELM Survey,][and references therein]{brown16a}, which
targets white dwarfs with $\log{g}<7$ and $M<0.3 M_{\odot}$. Given the finite
age of the universe, the only way to form ELM white dwarfs is through binary
evolution, and we do in fact find almost 100\% of ELM white dwarfs in short
period systems. This is significantly higher than the binary fraction of
10\% for the overall population of white dwarfs that were observed
as part of the Supernova-Ia Progenitor surveY \citep[SPY,][]{napiwotzki04,maoz17}. 

\citet{brown16b} estimate an ELM white dwarf merger rate of $3 \times 10^{-3}$
yr$^{-1}$ over the entire disk of the Milky Way. This is
significantly larger than the AM CVn formation rate, indicating that
most ELM white dwarf systems will merge. These merger
systems, depending on the total mass of the binary, will likely
form single subdwarfs, extreme helium stars, or single massive white dwarfs. The most
likely outcome is an R Cor Bor star, since the ELM white dwarf merger rate is
statistically identical to the R Cor Bor formation rate.
These merger rates are dominated by the quickest merger systems,
the ones with the shortest periods. 

There are currently five sub-hour orbital period detached double
white dwarfs known; J0106$-$1000, J0651+2844, J1630+4233, WD 0931+444
\citep[][and references therein]{kilic14}, and J1235+1543 \citep{breedt17}.
The two shortest period systems with $P<20$ min, J0651+2844 and WD 0931+444,
are verification sources for the {\em Laser Interferometer Space Antenna}
\citep[{\em LISA},][]{kilic15,korol17}. 
The discovery of additional sub-hour orbital period systems is
important for both precise white dwarf merger rate estimates and future space-based
gravitational wave missions \citep{amaro12}.

Here we present the results from a targeted search for sub-hour period
binary white dwarfs from Gemini Observatory, with additional follow-up observations
from the MMT. We discuss our target selection in Section 2, describe our
follow-up spectroscopy and photometry in Section 3, and present the orbital solutions
for four binaries in Section 4, including J1235+1543. \citet{breedt17}
independently identified the latter as a sub-hour orbital period system based on the SDSS
subspectra. Here we refine the orbital parameters of this system based on extended follow-up
observations. We discuss the parameters of the four confirmed binary systems in our sample, as well
as the implications of the results from this search in Section 5.

\section{Target Selection}

\begin{figure}
\includegraphics[width=3.0in,angle=0]{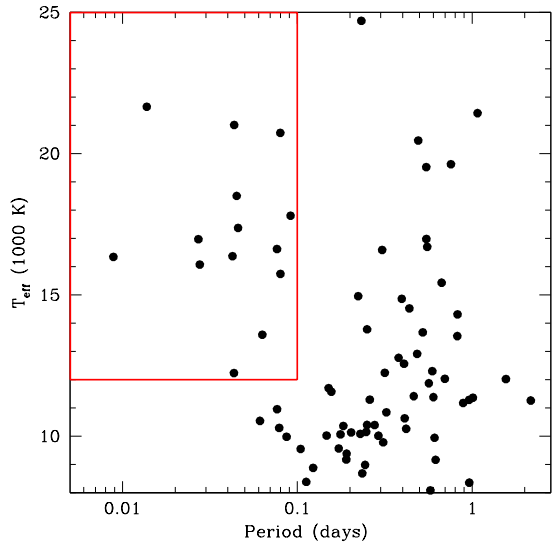}
\caption{Temperature versus orbital period distribution for the binary
white dwarfs in the ELM Survey. Roughly 39\% of the previously known
ELM white dwarfs with $T_{\rm eff}>12,000$ K are in binaries with $P<0.1$ d (red box).}
\label{fig:prev}
\end{figure}

Figure \ref{fig:prev} shows the temperature versus period distribution
of the binary white dwarfs in the ELM Survey \citep{brown16a}. This figure
demonstrates that the shortest period systems also happen to be
the hotter white dwarfs with $T_{\rm eff}>$ 12,000 K. This is a direct
consequence of gravitational wave emission: white dwarfs in the shortest period
systems merge before they have a chance to cool down. Hence, we only
see them when they are relatively young and hot. This provides an
excellent, but currently under-utilized, selection mechanism for the
shortest period binary systems. For example, 39\% of the previously
observed ELM white dwarfs hotter than 12,000 K are in binaries with $P<0.1$ d,
with a median period of 65 min.

We take advantage of this selection mechanism to search for short period
binary white dwarfs in the SDSS Data Release 10 spectroscopy sample. One of the authors (CAP)
fitted all of the DR10 optical spectra with stellar templates for
main-sequence stars \citep{castelli04} and white dwarfs using the {\tt FERRE} code \citep{allende15}.
Among these objects, we identify 49
relatively hot low-mass white dwarfs with $T_{\rm eff}>$ 12,000 K, $M<0.4 M_{\odot}$,
S/N$>$10 SDSS spectroscopy, and with no previous radial velocity observations. 
Note that our target selection would have included the eclipsing double white dwarf
systems J0651 \citep{brown11} and CSS 41177 \citep{parsons11}. 

\section{Observations}

We obtained follow-up optical spectroscopy of 34(10) targets using the 8m
Gemini North (South) telescope equipped with the Gemini Multi-Object
Spectrograph (GMOS) as part of the programs GN-2016A-Q-54, GN-2016B-Q-45,
GS-2015A-Q-10, GS-2016A-Q-58, and GS-2016B-Q-48. Since we are only interested
in finding sub-hour orbital period systems, and not constraining the binary
periods for all targets, we limited our observations to $\approx$30 min per
target. Depending on the target brightness, we obtained a sequence of
4-11 $\times$ 2-8 min long exposures with the B600 grating and a 
0.5$\arcsec$ slit, providing wavelength coverage from 3570 \AA\ to 6430 \AA\
and a resolving power of 1850 for GMOS-North, and coverage from
3620 \AA\ to 6780 \AA\ and a resolving power of 1940 for GMOS-South.
Each spectrum has a comparison lamp exposure taken within 10 min of
the observation time. 

Based on the initial velocity measurements from GMOS, we obtained 
additional follow-up data for six targets (J1113+2712, J1237+4913,
J1323+3254, J1407+1241, J1633+3030, and J1716+2838) using the same setup on
Gemini North as part of the Fast Turnaround program GN-2016A-FT-34.
Most of these targets were observed with back-to-back exposures over
$\approx$1.8 h, but some of the observations were split into multiple
nights due to weather conditions and the constraints imposed by queue scheduling.

We used the 6.5m MMT with the Blue Channel spectrograph to
obtain follow-up data on five targets (J0834+3049,
J1032+2147, J1235+1543, J1237+4913, and J2342+0811) between 2016 Jan and
2017 Mar. We operated the spectrograph with the 832 line mm$^{-1}$
grating in second order, providing wavelength coverage from
3600 \AA\ to 4500 \AA\ and a spectral resolution of 1.0 \AA. 
We obtained all observations at the parallactic angle, with a comparison
lamp exposure paired with every observation. We flux-calibrated using blue
spectrophotometric standards \citep{massey88}.

We also used the Kitt Peak National Observatory 4m telescope +
KOSMOS \citep{martini14} in 2016 Dec and the Apache Point Observatory 3.5m telescope
with the Dual Imaging Spectrograph (DIS) in 2017 Mar to obtain additional follow-up spectroscopy
of J1237+4913. We operated the KOSMOS (as part of the program
2016B-0160) and DIS spectrographs with the b2k and B1200 gratings, providing wavelength
coverages of 3500-6200 \AA\ and 3750-5000 \AA, and spectral resolutions of 2.0 \AA\ and
1.8 \AA, respectively.

We obtained follow-up time-series photometry of one of our targets,
J1235+1543, using the McDonald Observatory 2.1m Otto Struve telescope with the ProEM
camera and the BG40 filter. We used an exposure time of 10 s with a total
integration time of 3230 s, which covers the entire orbital period for this
short period system. We binned the CCD by $4\times4$, which resulted in a 
plate scale of $0.38\arcsec$ pixel$^{-1}$. We adopted the external IRAF package
{\em ccd\_hsp} \citep{kanaan02} for aperture photometry. There was only one
bright comparison star available in the field of view, and we corrected for
transparency variations by dividing the sky-subtracted light curve by
the light curve of this comparison star. 

\section{Results}

\subsection{Stellar Atmosphere Fits}

We employed a pure-hydrogen model atmosphere grid covering 4000-35,000 K
and $\log{g}=$ 4.5-9.5 to fit the normalized Balmer line profiles of
our targets in the summed, rest-frame Gemini spectra. The models
and our fitting procedures are described in \citet{gianninas15}.
We used the evoluationary sequences from \citet{althaus13} for 
low-mass He-core white dwarfs and \citet{fontaine01} for C/O core white dwarfs
to estimate masses and absolute magnitudes for each object.

\begin{figure}
\hspace{-0.1in}\includegraphics[width=3.7in,angle=0]{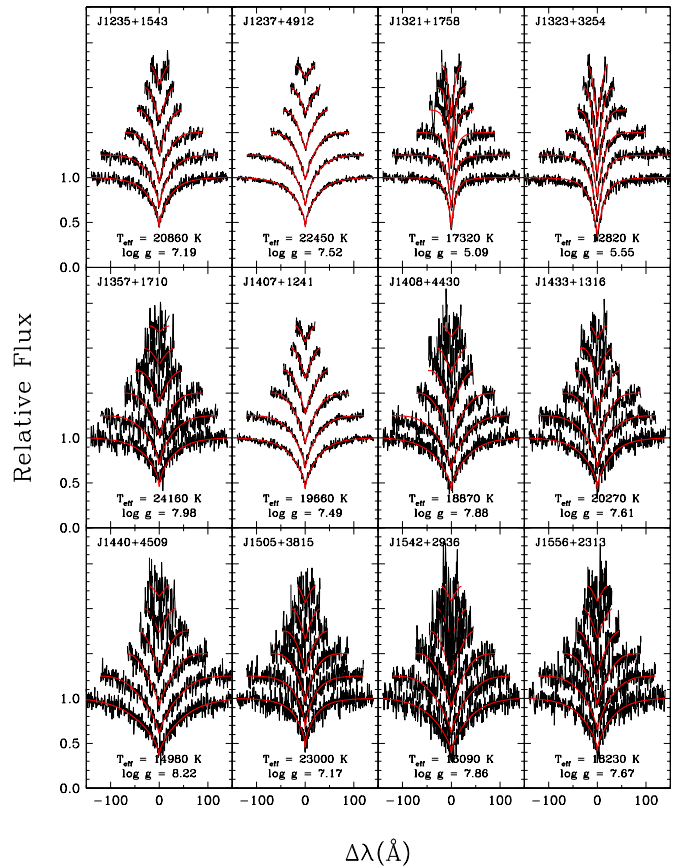}
\caption{1D model fits (red) to the observed Balmer line profiles (black)
from Gemini spectra for 12 of the targets in our sample. We fit the lines
from H$\beta$ (bottom) to H9 (top) using pure hydrogen model atmospheres.}
\label{fig:specfit}
\end{figure}

Figure \ref{fig:specfit} shows our model fits to a dozen targets in 
our sample. Given our initial target selection ($T_{\rm eff}\geq$ 12,000 K) 
based on the SDSS data, this figure uses the hot solution in the model fits.
Balmer lines are strongest at $T_{\rm eff} \sim$ 14,000 K
for average mass C/O white dwarfs. This usually leads to a degeneracy in the best-fit 
solution for model atmosphere analysis where a hot and a cool solution can both
fit the normalized Balmer line profiles reasonably well, but optical photometry
can help identify the correct solution in the majority of the cases. Five of
our targets have $ugriz$ photometry that implies an effective temperature below 10,000 K.
Using the cool solutions in our spectroscopic model fits, these five objects
(J1113+2712, J1321+1758, J1323+3254, J1011+0242, and J1132+0751) are best-fit
with $T_{\rm eff}\leq9,000$K and $\log{g}<7$ models, i.e. sdA stars \citep{kepler16}.
\citet{brown17} demonstrate that $\sim$99\% of the sdA stars 
are metal-poor main-sequence stars in the halo. Hence, we do not consider these five stars
as white dwarfs. 

\begin{table*}
\centering
\scriptsize
\caption{Physical parameters and the summary of radial velocity measurements for our targets. 
The top 34 and the bottom 10 objects were observed at Gemini North and South, respectively.}
\begin{tabular}{ccrcccrrc}
\hline
Object & $g_0$ & $T_{\rm eff}$ & $\log{g}$     & M            & $N$ & Mean & $\chi^2$ & $\log_{\rm 10}{(p)}$ \\ 
SDSS   & (mag) &  (K)          & (cm s$^{-2}$) & ($M_{\odot}$)&     & (\kms)&         &          \\ 
\hline
J073708.56+360215.0 & 19.08 & 23840 $\pm$ 630 & 7.50 $\pm$ 0.08 & 0.43 $\pm$ 0.03 & 6 & $ -76.6 \pm 28.3 $ & 18.93 & $ -2.70 $ \\
J073835.56+324121.6 & 19.55 & 23080 $\pm$ 730 & 7.69 $\pm$ 0.10 & 0.48 $\pm$ 0.05 & 6 & $ 10.4 \pm 30.7 $ & 32.81 & $ -5.39 $ \\
J074928.74+364451.6 & 19.06 & 24630 $\pm$ 710 & 7.89 $\pm$ 0.11 & 0.58 $\pm$ 0.06 & 6 & $ 67.5 \pm 37.2 $ & 7.93 & $ -0.80 $ \\
J080024.85+393757.6 & 19.08 & 25580 $\pm$ 480 & 7.84 $\pm$ 0.07 & 0.55 $\pm$ 0.04 & 6 & $ 25.9 \pm 27.7 $ & 14.86 & $ -1.96 $ \\
J082239.40+114142.7 & 18.75 & 25580 $\pm$ 460 & 7.45 $\pm$ 0.07 & 0.43 $\pm$ 0.03 & 5 & $ -30.9 \pm 21.1 $ & 5.98 & $ -0.70 $ \\
J083446.91+304959.2 & 18.78 & 17680 $\pm$ 380 & 7.06 $\pm$ 0.07 & 0.29 $\pm$ 0.01 & 18 & $ 210.4 \pm 7.5 $ & 228.98 & $ -38.4 $ \\
J090730.65+381754.0 & 17.83 & 22260 $\pm$ 280 & 7.25 $\pm$ 0.04 & 0.38 $\pm$ 0.01 & 8 & $ -69.3 \pm 15.0 $ & 26.10 & $ -3.32 $ \\
J091844.51+120948.1 & 17.92 & 18090 $\pm$ 190 & 7.39 $\pm$ 0.03 & 0.39 $\pm$ 0.02 & 7 & $ 83.9 \pm 13.9 $ & 3.18 & $ -0.10 $ \\
J103232.94+214712.3 & 18.21 & 15760 $\pm$ 260 & 7.47 $\pm$ 0.05 & 0.40 $\pm$ 0.02 & 29 & $ 24.6 \pm 5.0 $ & 58.06 & $ -3.14 $ \\
J111303.59+271259.0 & 18.53 & 8890 $\pm$ 290 & 5.56 $\pm$ 0.16 & \dots & 12 & $ 54.8 \pm 10.6 $ & 24.43 & $ -1.96 $ \\
J113117.50+374740.1 & 18.55 & 22600 $\pm$ 370 & 7.37 $\pm$ 0.05 & 0.40 $\pm$ 0.02 & 6 & $ -22.8 \pm 19.0 $ & 7.11 & $ -0.67 $ \\
J120020.71+682019.8 & 18.36 & 15360 $\pm$ 460 & 7.62 $\pm$ 0.07 & 0.44 $\pm$ 0.03 & 6 & $ 80.8 \pm 30.4 $ & 7.32 & $ -0.70 $ \\
J123549.88+154319.3 & 17.19 & 20860 $\pm$ 230 & 7.19 $\pm$ 0.03 & 0.35 $\pm$ 0.01 & 39 & $ -8.0 \pm 4.1 $ & 774.3 & $ -137 $ \\
J123728.64+491302.6 & 18.50 & 22450 $\pm$ 130 & 7.52 $\pm$ 0.02 & 0.43 $\pm$ 0.02 & 41 & $ -56.6 \pm 6.1 $ & 283.2 & $ -37.6 $ \\
J132153.51+175806.1 & 17.26 & 8070 $\pm$ 50 & 5.95 $\pm$ 0.17 & \dots  & 11 & $ 190.2 \pm 12.4 $ & 9.31 & $ -0.30 $ \\
J132350.96+325444.5 & 19.43 & 9100 $\pm$ 30 & 5.17 $\pm$ 0.16 & \dots  & 18 & $ -150.9 \pm 7.9 $ & 27.71 & $ -1.31 $ \\
J135715.36+171032.1 & 19.00 & 24160 $\pm$ 460 & 7.98 $\pm$ 0.06 & 0.62 $\pm$ 0.04 & 6 & $ 47.7 \pm 23.9 $ & 7.03 & $ -0.66 $ \\
J140714.50+124153.8 & 19.19 & 19660 $\pm$ 150 & 7.49 $\pm$ 0.02 & 0.42 $\pm$ 0.02 & 18 & $ -12.8 \pm 8.7 $ & 48.21 & $ -4.10 $ \\
J140821.99+443008.0 & 18.97 & 18870 $\pm$ 350 & 7.88 $\pm$ 0.06 & 0.56 $\pm$ 0.04 & 6 & $ -12.3 \pm 22.9 $ & 5.00 & $ -0.38 $ \\
J143315.47+131654.0 & 19.17 & 20270 $\pm$ 370 & 7.61 $\pm$ 0.05 & 0.45 $\pm$ 0.03 & 4 & $ -48.3 \pm 21.0 $ & 3.22 & $ -0.44 $ \\
J144023.92+450938.2 & 19.16 & 14980 $\pm$ 380 & 8.22 $\pm$ 0.05 & 0.75 $\pm$ 0.04 & 4 & $ 31.7 \pm 27.4 $ & 14.00 & $ -2.54 $ \\
J150546.21+381554.2 & 18.84 & 23000 $\pm$ 520 & 7.17 $\pm$ 0.07 & 0.36 $\pm$ 0.02 & 5 & $ -110.6 \pm 24.0 $ & 2.47 & $ -0.19 $ \\
J154230.67+293606.3 & 18.60 & 16090 $\pm$ 450 & 7.86 $\pm$ 0.08 & 0.54 $\pm$ 0.05 & 5 & $ 15.6 \pm 29.2 $ & 1.97 & $ -0.13 $ \\
J155657.69+231358.5 & 18.68 & 18230 $\pm$ 380 & 7.67 $\pm$ 0.06 & 0.46 $\pm$ 0.03 & 4 & $ 25.4 \pm 19.1 $ & 8.41 & $ -1.42 $ \\
J163338.88+303041.7 & 18.16 & 18740 $\pm$ 110 & 7.30 $\pm$ 0.02 & 0.37 $\pm$ 0.01 & 24 & $ -120.7 \pm 6.9 $ & 35.26 & $ -1.31 $ \\
J170816.36+222551.0 & 19.07 & 22900 $\pm$ 770 & 7.14 $\pm$ 0.10 & 0.34 $\pm$ 0.03 & 4 & $ -52.6 \pm 26.6 $ & 10.20 & $ -1.77 $ \\
J171602.17+283852.3 & 18.94 & 17700 $\pm$ 120 & 7.69 $\pm$ 0.02 & 0.46 $\pm$ 0.02 & 18 & $ -7.2 \pm 9.3 $ & 37.84 & $ -2.59 $ \\
J172850.05+581316.3 & 18.52 & 23790 $\pm$ 610 & 7.51 $\pm$ 0.08 & 0.43 $\pm$ 0.03 & 5 & $ -61.9 \pm 22.9 $ & 1.93 & $ -0.13 $ \\
J201154.21$-$104124.1 & 18.90 & 22530 $\pm$ 770 & 7.95 $\pm$ 0.10 & 0.60 $\pm$ 0.06  & 5 & $ 73.1 \pm 31.3 $ & 10.99 & $ -1.57 $ \\
J210252.10+010108.6 & 18.20 & 22980 $\pm$ 270 & 7.59 $\pm$ 0.04 & 0.45 $\pm$ 0.02 & 5 & $ 105.1 \pm 16.3 $ & 3.85 & $ -0.37 $ \\
J221426.78+055025.8 & 18.66 & 22090 $\pm$ 450 & 7.22 $\pm$ 0.06 & 0.37 $\pm$ 0.02 & 5 & $ -46.2 \pm 19.0 $ & 24.76 & $ -4.25 $ \\
J224750.14+295145.1 & 19.25 & 22150 $\pm$ 480 & 8.05 $\pm$ 0.06 & 0.66 $\pm$ 0.04 & 5 & $ -39.2 \pm 28.2 $ & 29.20 & $ -5.15 $ \\
J234212.47+005121.0 & 19.30 & 17230 $\pm$ 340 & 7.53 $\pm$ 0.06 & 0.42 $\pm$ 0.03 & 5 & $ -5.3 \pm 19.8 $ & 6.87 & $ -0.84 $ \\
J234248.86+081137.2 & 18.32 & 22030 $\pm$ 230 & 7.45 $\pm$ 0.03 & 0.42 $\pm$ 0.02 & 25 & $ 12.5 \pm 5.7 $ & 239.3 & $ -36.7 $ \\
\hline
J000437.66$-$055731.4 & 19.03 & 14220 $\pm$ 210 & 7.76 $\pm$ 0.03 & 0.48 $\pm$ 0.03 & 5 & $ -30.8 \pm 14.8 $ & 8.53 & $ -1.13 $ \\
J011258.36$-$005952.4 & 18.57 & 18720 $\pm$ 120 & 7.55 $\pm$ 0.02 & 0.43 $\pm$ 0.02 & 5 & $ 53.3 \pm 12.1 $ & 1.98 & $ -0.13 $ \\
J031504.58$-$065727.2 & 17.45 & 19120 $\pm$ 120 & 7.48 $\pm$ 0.02 & 0.41 $\pm$ 0.02 & 9 & $ 61.0 \pm 11.3 $ & 5.75 & $ -0.17 $ \\
J091911.72+082004.4 & 19.26 & 25040 $\pm$ 360 & 7.29 $\pm$ 0.05 & 0.39 $\pm$ 0.02 & 6 & $ 27.2 \pm 16.3 $ & 8.27 & $ -0.85 $ \\
J101132.73+024216.4 & 18.81 & 8900 $\pm$ 70 & 6.50 $\pm$ 0.16 &  \dots & 4 & $ 325.1 \pm 29.9 $ & 0.26 & $ -0.01 $ \\
J103702.02+032648.0 & 18.29 & 23400 $\pm$ 400 & 7.25 $\pm$ 0.05 & 0.38 $\pm$ 0.02 & 8 & $ 109.3 \pm 19.2 $ & 14.67 & $ -1.39 $ \\
J113218.41+075103.0 & 16.94 & 7010 $\pm$ 30 & 6.78 $\pm$ 0.07 &  \dots  & 9 & $ 37.9 \pm 14.0 $ & 4.60 & $ -0.10 $ \\
J123717.06$-$003900.1 & 19.46 & 16020 $\pm$ 350 & 8.02 $\pm$ 0.06 & 0.63 $\pm$ 0.04 & 6 & $ 71.0 \pm 29.0 $ & 1.54 & $ -0.04 $ \\
J154647.28$-$005003.5 & 17.68 & 21410 $\pm$ 430 & 7.46 $\pm$ 0.06 & 0.42 $\pm$ 0.02 & 8 & $ 51.7 \pm 18.3 $ & 16.19 & $ -1.63 $ \\
J162024.40$-$000545.9 & 17.03 & 21160 $\pm$ 260 & 7.43 $\pm$ 0.04 & 0.41 $\pm$ 0.02 & 13 & $ 30.7 \pm 14.2 $ & 23.68 & $ -1.65 $ \\
\hline
\end{tabular}
\end{table*}

Table 1 presents the physical parameters for all 44 stars in our sample.
Our model atmosphere analysis shows that nine of these stars have masses above
$0.5 M_{\odot}$. Excluding these nine
stars and the sdAs, there are 30 low-mass white dwarfs in our sample.

\subsection{Radial Velocities and Errors}

        We measure radial velocities by cross-correlating the spectra against
high signal-to-noise templates of known velocity.  We use the RVSAO package
documented in \citet{kurtz98} and based on the \citet{tonry79} algorithm.  The
cross-correlation is the normal product of the Fourier transform of an object
spectrum with the conjugate of the Fourier transform of a template spectrum.
The software package includes extra steps such as Fourier bandpass filtering,
to dampen the high frequency (pixel-to-pixel) and low frequency (slow continuum
roll) noise in spectra.

        Velocity errors are measured from the full-width-at-half-maximum of the
cross-correlation peak using the $r$-statistic \citep{tonry79}.  Empirical
validation using repeat low- and high-spectral resolution observations of
galaxies and stars confirm the precision of the cross-correlation errors
\citep{kurtz98}. However, systematic errors can arise from poorly-matched
templates, which skew the shape of the cross-correlation peak.

        We address this systematic issue by shifting-to-rest-frame and summing
together all observations of a given target, and then cross-correlating the
individual spectra against the summed spectrum of itself.  This approach
minimizes systematic error, but hides statistical error.  The location of a
star on the spectrograph slit changes how it illuminates the spectrograph and
disperses its light onto the detector.  We use our higher resolution and higher signal-to-noise
ratio MMT data to investigate this issue. Back-to-back exposures of constant
velocity targets at the MMT demonstrate a 10 to 15 \kms\ dispersion slit
illumination effect, most apparent in targets observed in sub-arcsec seeing
with short exposure times (like J1235+1543).  Wavelength calibration errors also
contribute.  The MMT arc line fits have 3 \kms\ residuals, however the blue end
$<$3900 \AA\ is anchored by weak lines that have larger 5 to 10 \kms\
residuals.  The cross-correlation does not discriminate between slit
illumination and wavelength calibration errors and real velocity change.  The
upshot is that we must add statistical error in quadrature to the
cross-correlation error.

        Our approach is to add 20 \kms\ in quadrature to the velocity errors of
objects observed at Gemini and MMT and 30 \kms\ in quadrature to the velocity
errors of objects observed at KPNO and APO. When fitting orbital parameters to
the confirmed binaries, this choice of errors yields reduced $\chi^2$ values of 1
(see the discussion in Section 4.4). We also test for zero point offsets between
telescopes when fitting binary orbital parameters.  We see no evidence for zero
point offsets greater than the 1-$\sigma$ error in $\gamma$, the systemic velocity.

\subsection{Constraints on Radial Velocity Variability}

\citet{maxted00} presented a robust method for identifying radial velocity variable
objects. They used the weighted mean radial velocity for each star in their sample
to calculate the $\chi^2$ statistic for a constant-velocity model. They then calculated 
the probability, $p$, of obtaining the observed value of $\chi^2$ or higher from random
fluctuations of a constant velocity, taking into account the appropriate number of
degrees of freedom. They identify objects with $\log{(p)}<-4$ as binary systems. This
selection leads to a false detection rate of $<0.5$\% in a sample of 44 objects. 

We adopt the same method to identify radial velocity variable objects
in our sample. Table 1 lists the number of spectra ($N$), weighted mean radial velocity,
$\chi^2$ for a constant-velocity fit, and the probability of obtaining this $\chi^2$
value given the number of degrees of freedom ($N-1$) for each star. The majority of
the stars in our sample do not show significant velocity variations. Given the brevity
of our Gemini observations (except for the stars with extensive follow-up observations),
this is not surprising.

\begin{figure}
\includegraphics[width=3.3in,angle=0]{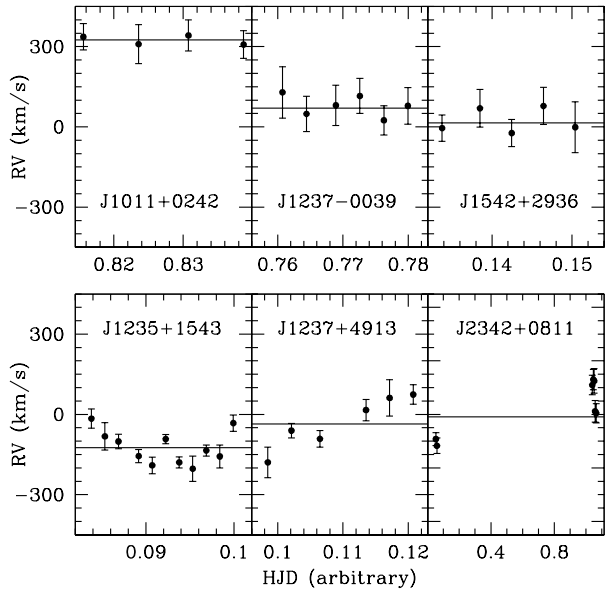}
\caption{Radial velocity measurements for 3 non-variable (top panels)
and 3 variable (bottom panels) objects observed at Gemini North and South.
The solid line shows the average velocity for each star.}
\label{fig:vnv}
\end{figure}

Figure \ref{fig:vnv} shows our initial set of Gemini observations for six stars,
three of which are excellent examples of non-velocity-variable objects. Our Gemini data
for J1011+0242, J1237$-$0039, and J1542+2936 (top panels) are consistent with a constant velocity fit
with $\log{(p)}=-0.01$ to $-0.13$. On the other hand, there are several targets with
$\log{(p)}<-4$, indicating that they are likely in short period binary systems (bottom panels).
For example, our 2 min cadence data on J1235+1543 sample a significant
portion of the binary orbit, and our 8 min cadence data on J1237+4913 reveal a positive
velocity trend in that system. Similarly, the initial set of Gemini observations on
J2342+0811 reveal a $\approx$250 \kms\ velocity change over two consecutive nights.
We discuss these three objects further in the next section.

The five sdA stars in our sample (J1113+2712, J1321+1758, J1323+3254,
J1011+0242, and J1132+0751) do not show significant velocity variations in our
data. Metal-poor main sequence stars in detached binaries must have orbital
periods above about 9 hr \citep{brown17}. Hence, the lack of significant velocity
variations in these stars, as well as the majority of the stars in our sample of low-mass
white dwarfs is consistent with the expectation that they are likely in longer period
binary systems.

\subsection{Four Binary Systems}

There are eight objects in Table 1 with $\log{(p)}<-4$; a constant velocity
model is a poor representation of the data for these stars. These are likely
binary systems. We have limited follow-up data on four of them, J0738+3241,
J1407+1241, J2214+0550, and J2247+2951, and we are unable to constrain the
orbital parameters for these four systems. However, the remaining four stars
with $\log{(p)}<-4$ have extensive follow-up observations, and they do show
significant radial velocity variations with periods ranging from 53 min to 7 h.

\begin{figure*}
\includegraphics[width=3.0in]{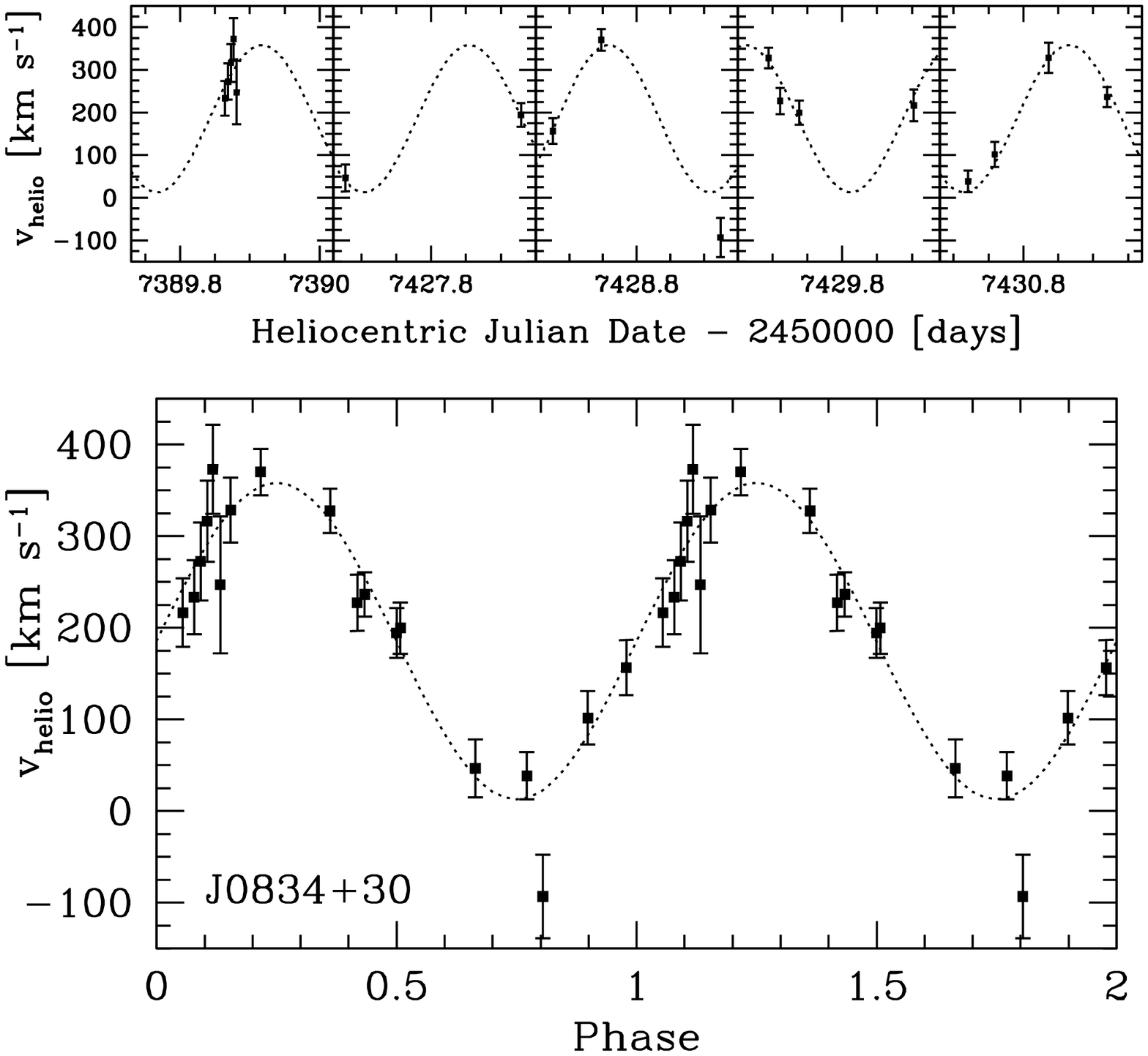}
\includegraphics[width=3.0in]{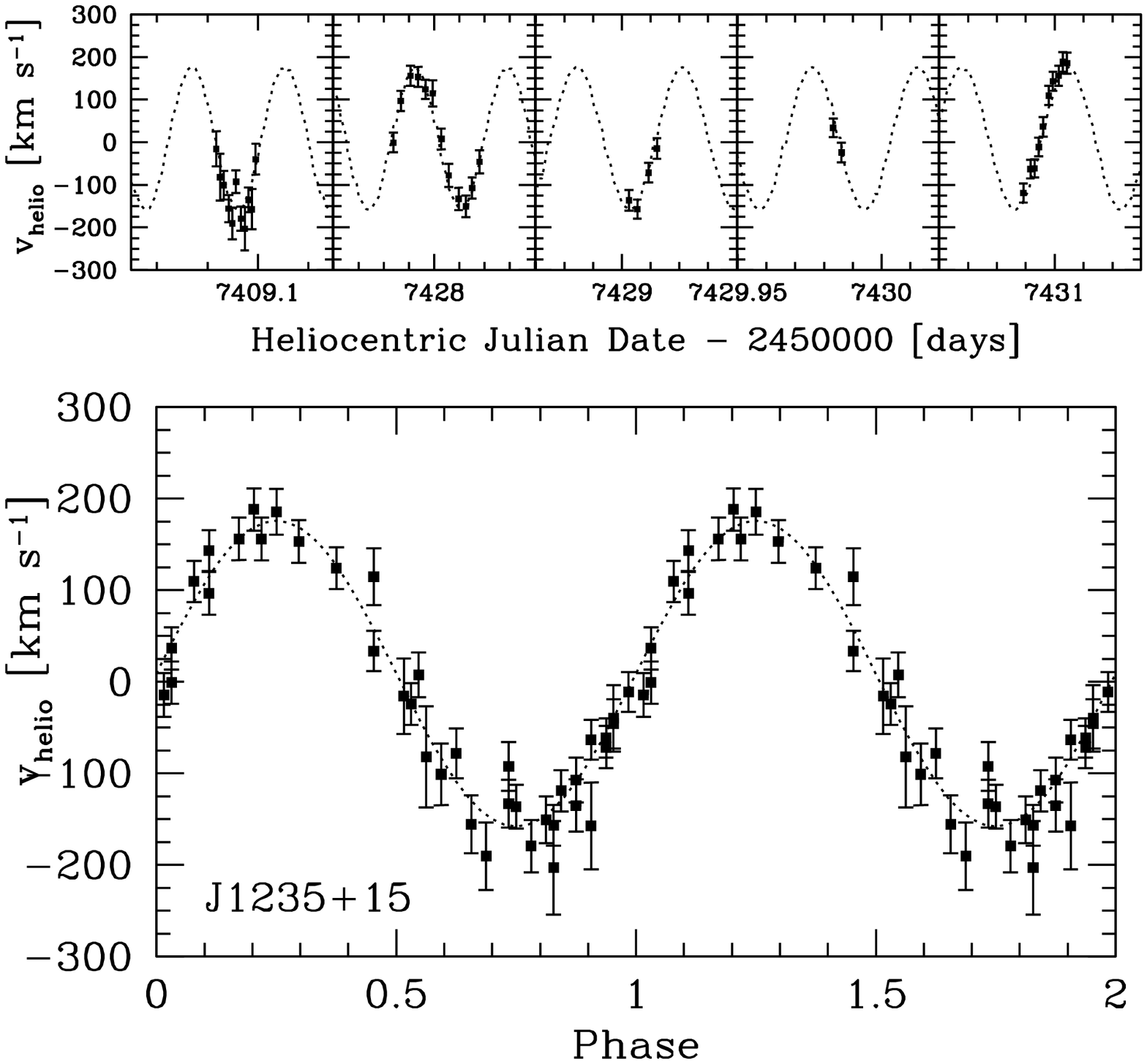}\\
\includegraphics[width=3.0in]{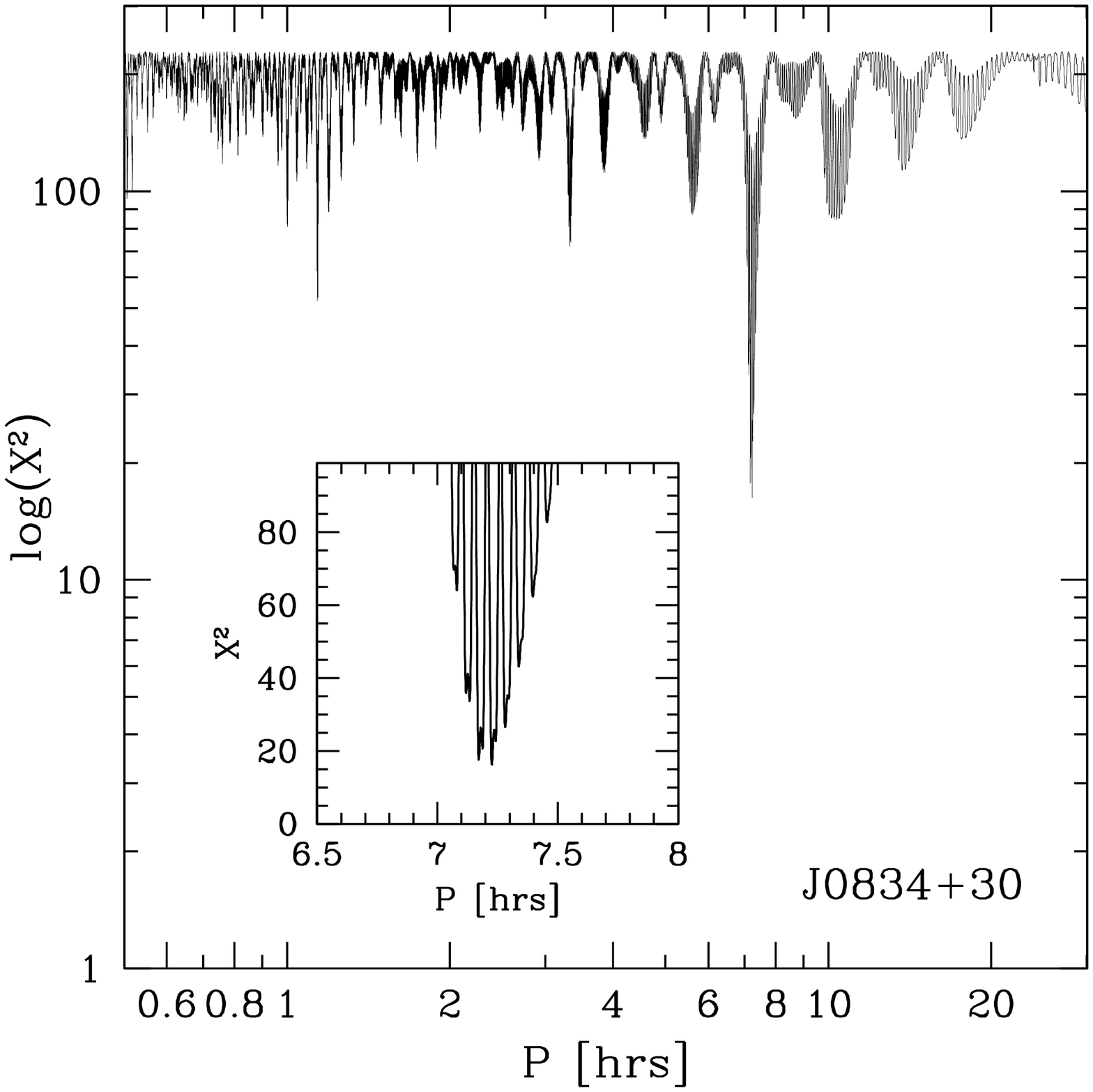}
\includegraphics[width=3.0in]{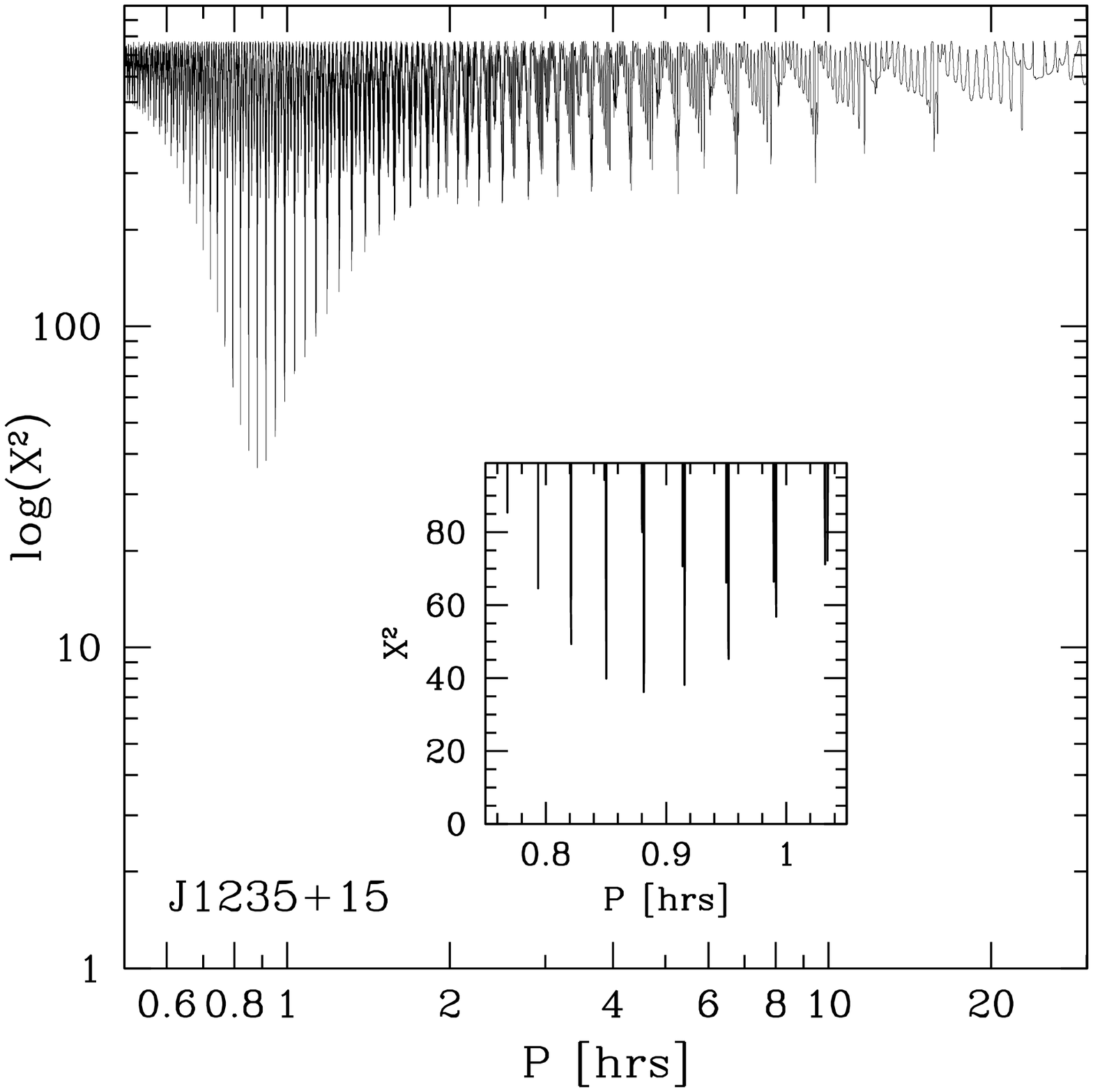}
\caption{Radial velocity observations, phased velocity curves, and
periodograms for J0834+3049 and J1235+1543. Insets show the
distribution of $\chi^2$ around the minima on a linear scale.}
\label{fig:bin1}
\end{figure*}

\begin{figure*}
\includegraphics[width=3.0in]{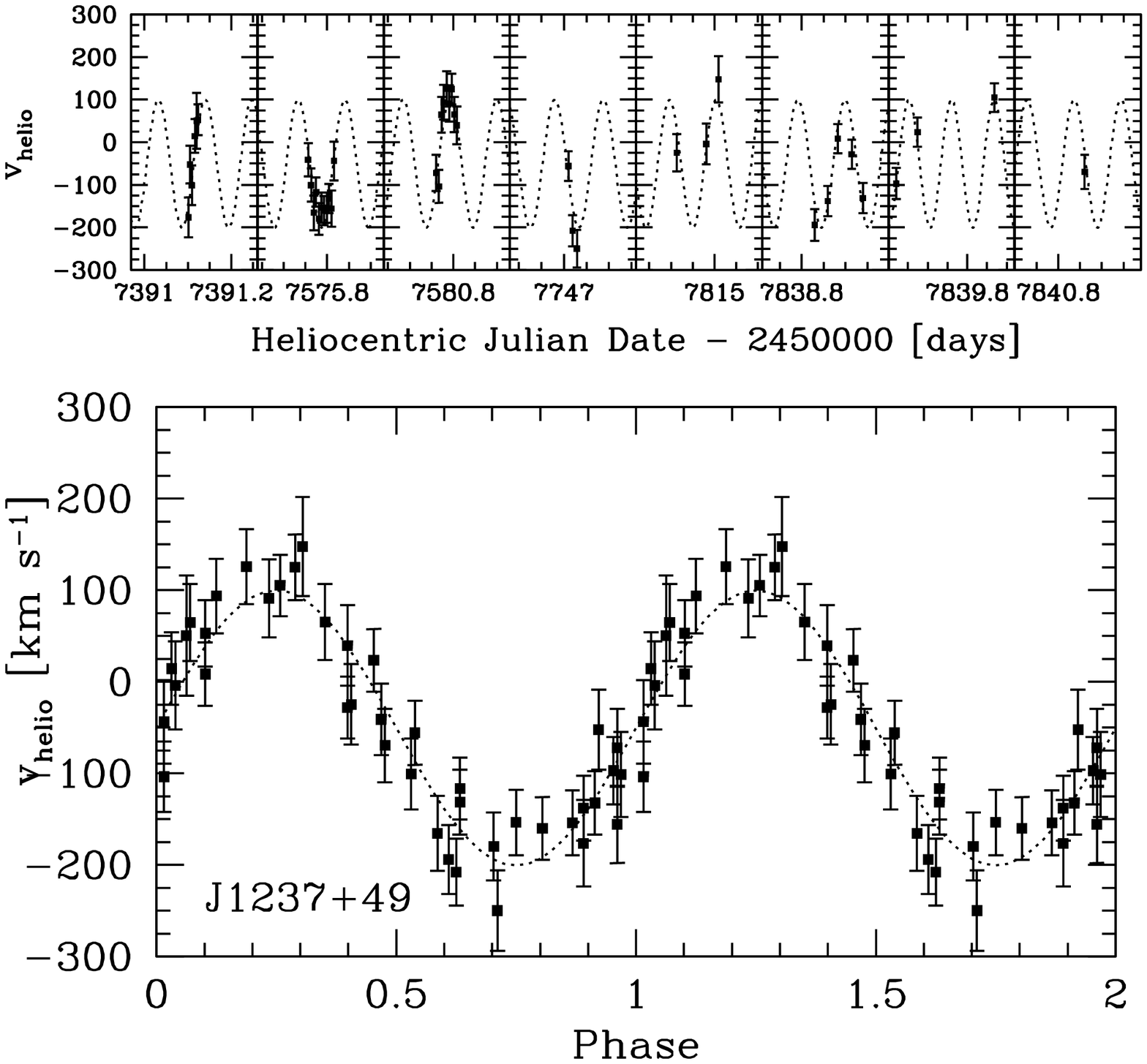}
\includegraphics[width=3.0in]{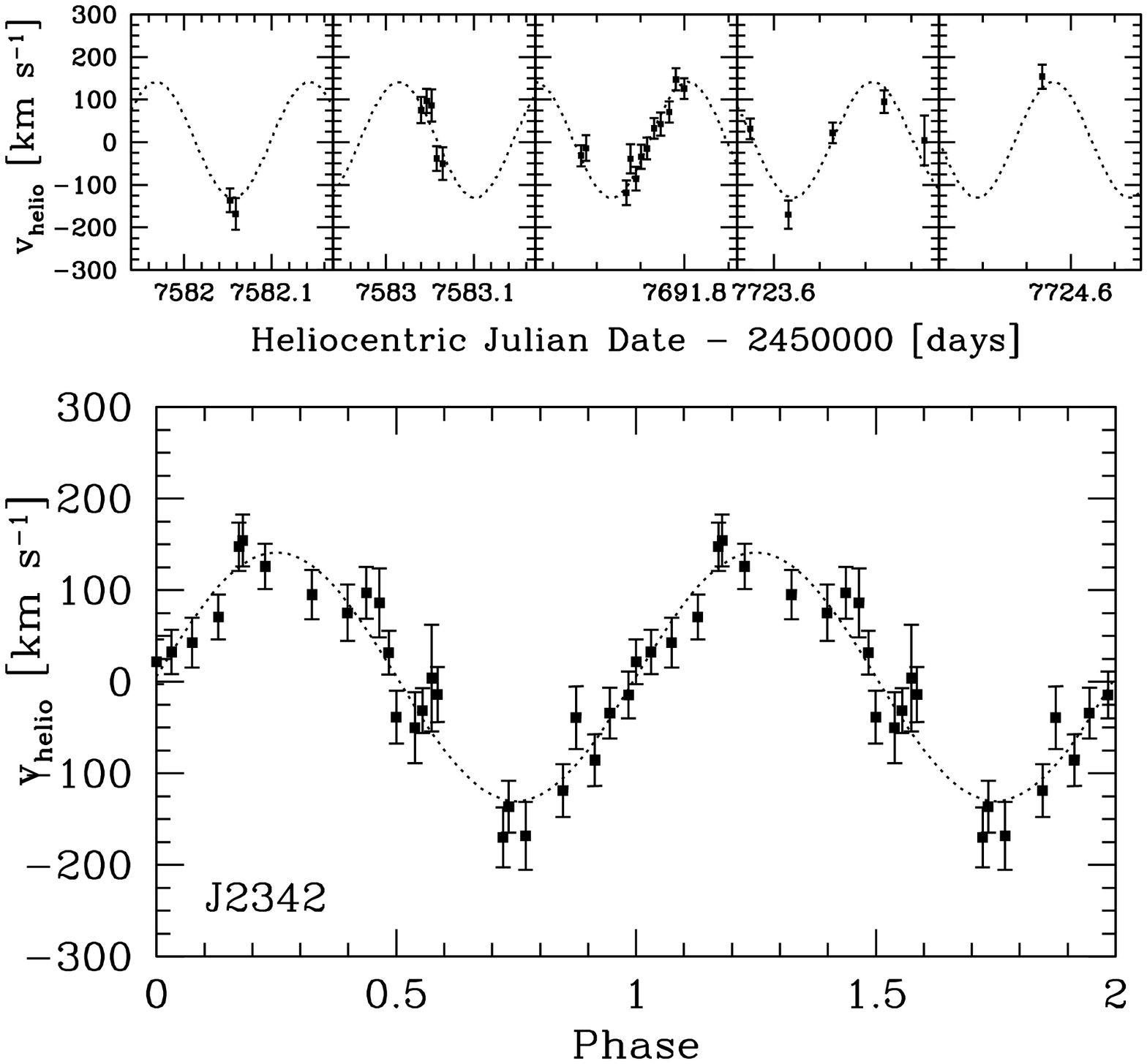}\\
\includegraphics[width=3.0in]{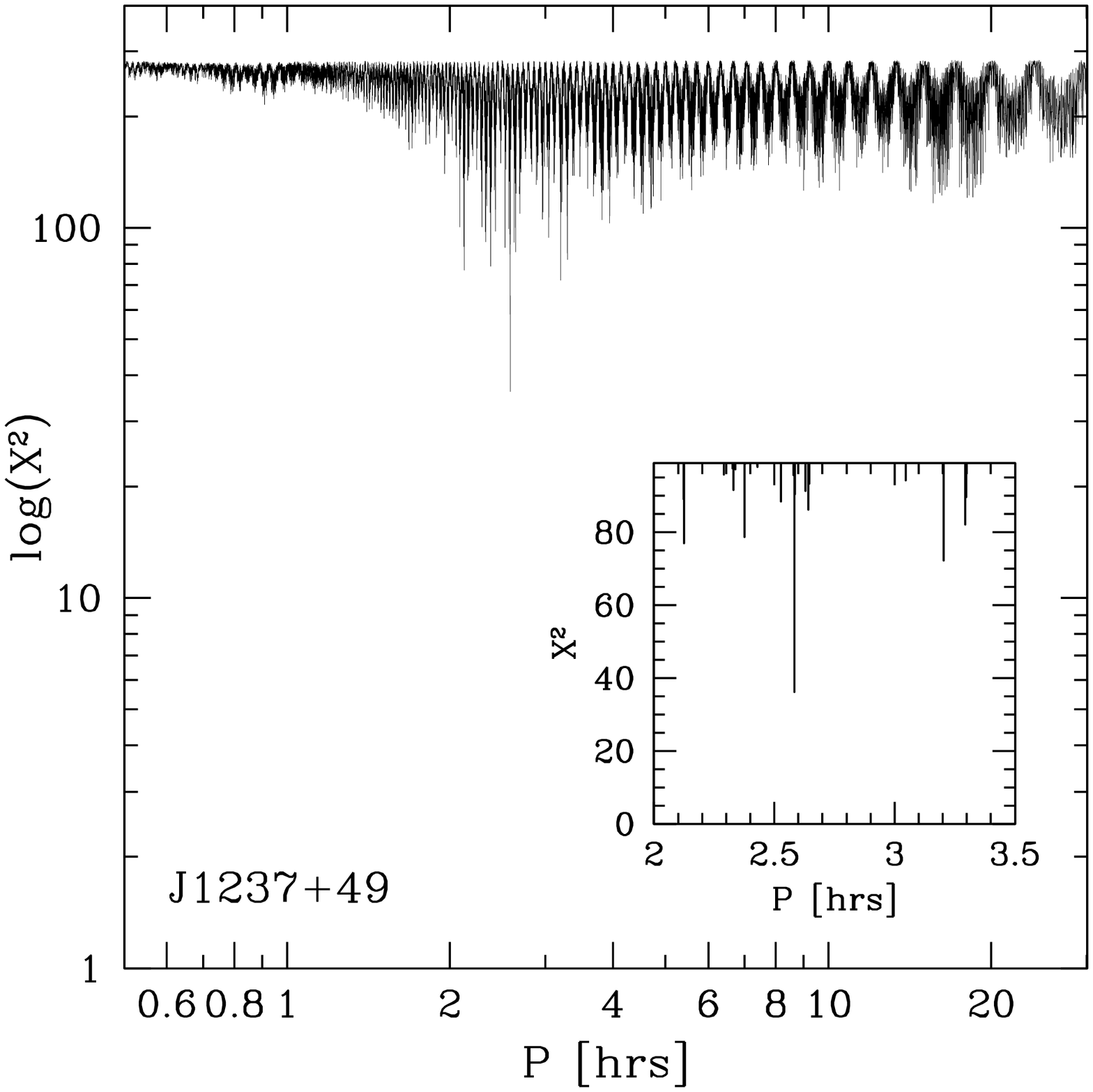}
\includegraphics[width=3.0in]{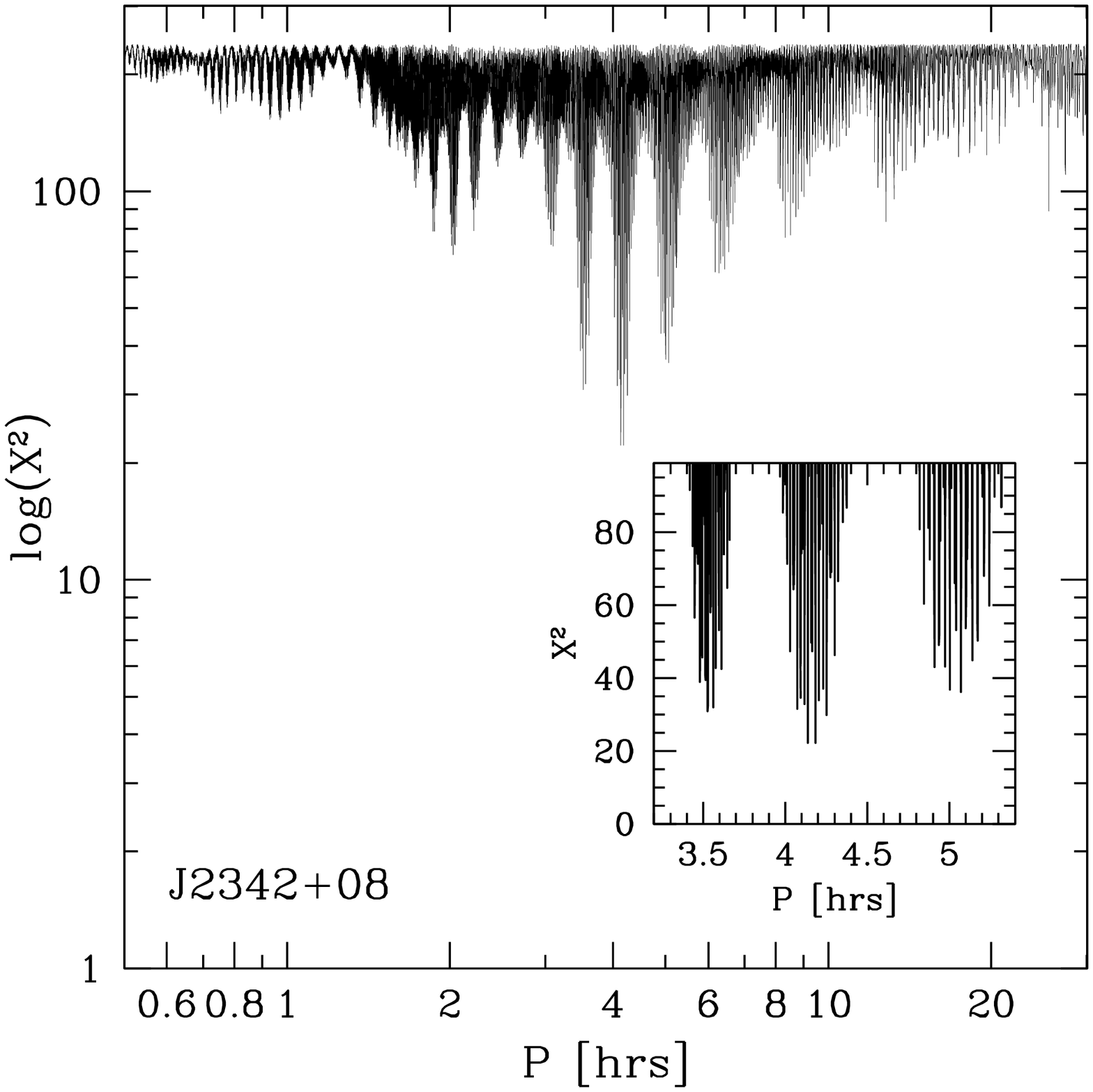}
\caption{Same as Figure \ref{fig:bin1}, but for J1237+4913 and J2342+0811.}
\label{fig:bin2}
\end{figure*}

We determine orbital parameters by minimizing $\chi^2$ for a circular orbit.
Figures \ref{fig:bin1} and \ref{fig:bin2} show the radial velocity
observations, phased velocity curves, and periodograms for these four white dwarfs. 
Each panel also includes a blow-up of the frequency range where the minimum
$\chi^2$ is found. \citet{morales03} discussed the problems with identifying the correct
orbital period from radial velocity data given problems with aliasing. They found
that the reduced $\chi^2$ values from circular orbit fits were significantly larger
than 1 for some of their targets. They attributed this to an unaccounted source of error
in their velocity measurements, perhaps the true variability of the star or slit illumination
effects. They estimated the level of this uncertainty in their data such that
when systematic and statistical errors are added in quadrature they give reduced
$\chi^2$ values of 1. We estimate statistical uncertainties of 20 and 30 \kms\ for the Gemini/MMT
and KPNO/APO data, respectively (Section 4.2). Adding the cross-correlation errors from
RVSAO and the statistical uncertainties in quadrature, we find the best-fit circular orbits
with reduced $\chi^2$ ranging from 0.97 to 1.16 for our four binary systems.

Out of these four binaries, three have unique orbital period solutions, while 
J2342+0811 has a significant period alias (at $P = 0.14369 \pm 0.0029$ d and
$K = 126.0 \pm  10.1$ \kms). Its second period alias at $P=5$ h differs by 20
in $\chi^2$ and is unlikely to be significant. The $\chi^2$ minima have substructure due to the
sampling (see the insets in Figures \ref{fig:bin1} and \ref{fig:bin2}), however we do not fit the
substructure.  We measure the orbital period from the envelope of $\chi^2$, which
is well-defined and symmetric in all four binaries.

We estimate errors by re-sampling the radial velocities with their
errors and re-fitting orbital parameters 10,000 times. This Monte Carlo
approach samples $\chi^2$ space in a self-consistent way. We report the median
period, semi-amplitude, and systemic velocity along with the average
15.9\% and 84.1\% percentiles of the distributions in Table \ref{table:orbit}.  The
distributions are symmetric, and so we average the percentiles for simplicity.
We also fit J2342+0811's second and third minima at $P=3.5$ and 5 hr;
the semi-amplitudes differ by 2 \kms\ and are thus statistically identical
to the best fit.

\begin{table*}
\centering
\caption{Orbital Parameters}
\begin{tabular}{ccccccccc}
\hline
SDSS & $P$ & $K$    & $\gamma$ & $\chi^2_{\rm red}$ & $f$           & $M_1$ & $M_2$ & $\tau_{\rm merge}$ \\
     & (d) & (\kms) & (\kms)   &                    & ($M_{\odot}$) & ($M_{\odot}$) & ($M_{\odot}$) & \\
\hline
J0834+3049 & 0.30079 $\pm$ 0.0011 & 179.3 $\pm$ 13.9 &  183.3 $\pm$ 8.5 & 1.16 & 0.1789 & 0.29 & $\geq$0.47 & $\leq$13 Gyr \\
J1235+1543 & 0.03672 $\pm$ 0.0014 & 166.5 $\pm$  6.2 &   10.5 $\pm$ 5.0 & 1.03 & 0.0176 & 0.35 & $\geq$0.17 & $\leq$98 Myr \\
J1237+4913 & 0.10763 $\pm$ 0.0024 & 143.6 $\pm$ 10.5 & $-45.1 \pm  7.0$ & 0.97 & 0.0334 & 0.43 & $\geq$0.25 & $\leq$1.0 Gyr \\
J2342+0811 & 0.16788 $\pm$ 0.0014 & 128.3 $\pm$ 10.9 &   9.1 $\pm$ 10.4 & 1.06 & 0.0367 & 0.42 & $\geq$0.26 & $\leq$3.3 Gyr \\
\hline
\end{tabular}
\label{table:orbit}
\end{table*}

Table 2 presents the orbital elements for these four binary systems with well constrained
orbits. Note that
\citet{breedt17} also identified J1235+1543 as a subhour orbital period binary
with $P=49.5$ min and $K=176 \pm 21$ km s$^{-1}$. However, they only used 5 radial velocity
measurements from 800-1000 s long exposures. Based on 39 exposures, with exposure times as short as
2 min, we refine the period and velocity semi-amplitude for J1235+1543 to
$P=$ 52.9 min  and $K=166.5 \pm 6.2$ \kms.

The observed velocity semi-amplitudes are relatively modest ($K<200$ \kms) for
these stars, even for the 53 min period system J1235+1543. The median semi-amplitude
of the ELM white dwarf binaries is 220 \kms \citep{brown16b}. However, our targets
are about twice as massive as the typical ELM white dwarfs, hence
the observed smaller velocity amplitudes are not surprising.

Table 2 also presents the mass functions, constraints on the companion masses,
and the merger times due to gravitational wave radiation. Note that we define the
visible low-mass white dwarf in each system as the primary star. The minimum mass companions
to our targets range from 0.17 to 0.47 $M_{\odot}$, with gravitational wave merger
times of roughly 100 Myr for J1235+1543 to $\leq$13 Gyr for J0834+3049. 
All but one of these objects, J0834+3049, have minimum mass companions that are smaller in mass than
the visible white dwarfs. Since lower mass white dwarfs should form last, and hence appear brighter,
these three single-lined spectroscopic binary systems are likely low inclination systems
where the companions are more massive than the visible white dwarfs.

\subsection{Photometric Constraints}

Spectral types and temperatures of the companions to single-lined spectroscopic
binaries can be inferred through photometric effects like Doppler boosting,
ellipsoidal variations, and eclipses, or through excess flux in the red or infrared bands. 
Based on our model atmosphere analysis, the absolute magnitudes of our four
binary white dwarfs range from 9.9 to 10.4 in the $i$-band. If the companions are M
dwarfs, the minimum mass companions would be comparable in brightness
\citep[within a factor of two,][]{kroupa97} or even brighter than our white dwarf
targets in the $i$-band. We do not see that. Hence, these four binary
systems are double degenerates.

\begin{figure}
\includegraphics[width=3.3in,angle=0]{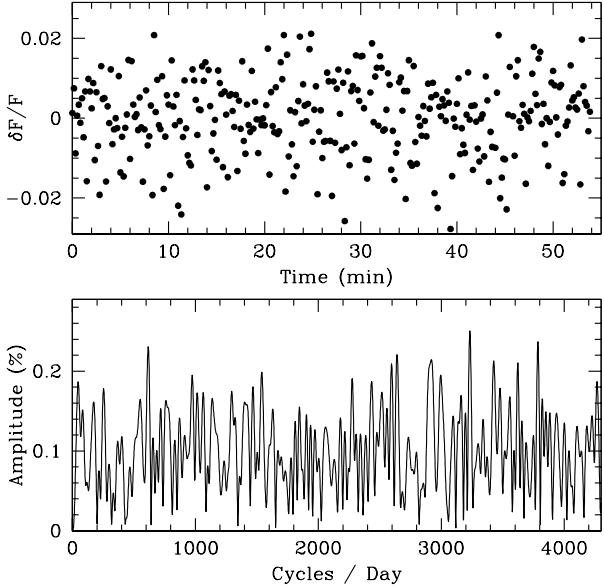}
\caption{High speed photometry of J1235+1543 over 54 min (top panel) and its
Fourier Transform (bottom panel). This short period binary does not show
any significant variability.}
\label{fig:1235}
\end{figure}

The probability of eclipses increases with decreasing orbital period. To search
for eclipses and other photometric effects, we obtained high-speed photometry of
the shortest period system in our sample, J1235+1543, with a cadence of 10 s.  
Figure \ref{fig:1235} shows these observations over a binary orbit. There is no
significant variability in this system, ruling out eclipses and ellipsoidal variations.
The amplitude of the ellipsoidal effect is proportional to
$(M_2/M_1)(R_1/a)^3$, where $a$ is the orbital semi-major axis and $R_1$
is the radius of the primary \citep{shporer10}. Compared to the ELM white dwarfs
that show ellipsoidal variations \citep{hermes14,bell17}, $(M_2/M_1)$ and $R_1$ are
relatively small for J1235+1543. Therefore, the
lack of ellipsoidal variations is not surprising.

\begin{figure}
\includegraphics[width=3.3in,angle=0]{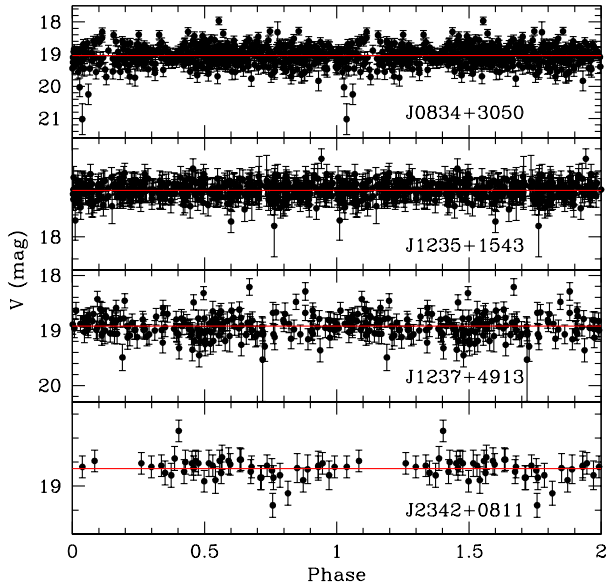}
\caption{Catalina Sky Survey light curves for the four binary white dwarfs in
our sample.}
\label{fig:crts}
\end{figure}

All four of the binary white dwarfs in our sample were observed by the Catalina
Sky Survey \citep{drake09}. Figure \ref{fig:crts} shows these light curves
phased with the best-fit period from the radial velocity data. The Catalina
data are sparse for J2342+0811 and part of the orbit is not covered. In
addition, the data are noisy for these relatively faint stars. There is a
$4\sigma$ dip in the J0834+3050 light curve that might be an eclipse,
however there are several other $>4\sigma$ outliers in the same light curve.
We suspect that the photometric errors are underestimated. We conclude that
there is no significant evidence for eclipses or other photometric effects
in any of these systems given the Catalina observations.

\section{Discussion}

\begin{figure}
\includegraphics[width=3.3in,angle=0]{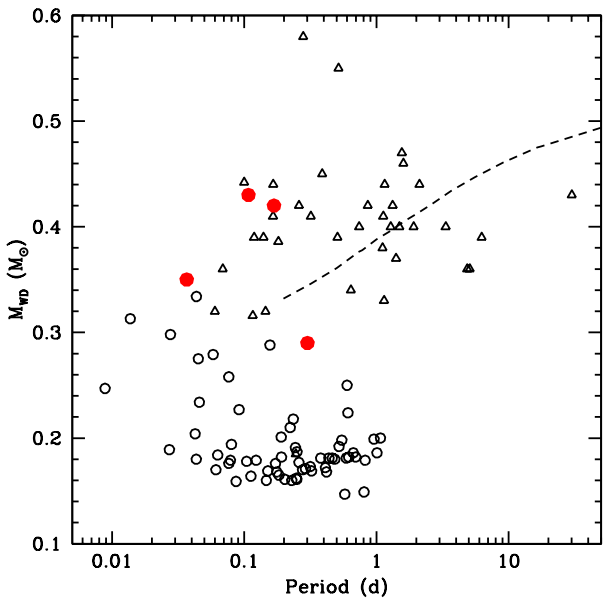}
\caption{Mass versus period distribution for ELM white dwarfs (open circles),
low-mass white dwarfs (open triangles), and the four binary systems
presented in this paper (filled circles). The dashed
line shows the results from the rapid binary-star evolution (BSE) algorithm of
\citet{hurley02} for a $2 M_{\odot} + 1 M_{\odot}$ main-sequence binary.}
\label{fig:mp}
\end{figure}

Our snapshot radial velocity survey of relatively hot and young low-mass white dwarfs
has revelaed four double degenerates with periods ranging from 53 min to about 7 h.
Figure \ref{fig:mp} compares the mass and period distribution for these systems against
the period distribution of ELM \citep{brown16b} and low-mass white dwarfs
\citep{nelemans05,brown11,debes15,hallakoun16,breedt17,rebassa17}. The dashed line
shows the predicted mass (of the brighter white dwarf) versus period relation from
the rapid binary-star evolution (BSE) algorithm of \citet{hurley02} for an
initial binary of main-sequence stars with masses $2 M_{\odot} + 1 M_{\odot}$.

The BSE calculations depend on two important parameters, $\alpha_{\rm CE}$ and
$\alpha_{\rm int}$ (or $\alpha_{\rm rec}$). The former parameter is the efficiency
in converting orbital energy into kinetic energy to eject the envelope, and the latter describes
the fraction of the internal energy (thermal, radiation and recombination energy)
used to eject the envelope. Note that the latest version of the BSE code treats the
binding energy parameter $\lambda$ as a variable. 
\citet{zorotovic10,zorotovic14}, \citet{toonen13}, and
\citet{cojocaru17} demonstrate that both of these efficiency parameters are small.
We adopt $\alpha_{\rm CE} = \alpha_{\rm rec} = 0.25$ as in \citet{zorotovic10,zorotovic14}
for the evolutionary sequence shown in Figure \ref{fig:mp}.
The BSE calculations demonstrate that the closest stellar pairs that survive
the common-envelope evolution should form lower mass white dwarfs. 
This is also consistent with the binary population synthesis calculations of
\citet[][see their Figure 8]{nelemans05}.

Studying the orbital period
distribution of post-common-envelope binaries containing C/O and He-core white dwarfs
separately, \citet{zorotovic11} found median periods of 0.57 d and 0.28 d for the two
samples respectively. This difference is consistent with our understanding of
the common-envelope evolution. If the mass
transfer starts when the primary star is on
the red giant branch, this leads to a He-core white dwarf, whereas
if the mass transfer starts while the primary is on the asymptotic giant branch, this
leads to a C/O core white dwarf. Hence, stellar evolution theory predicts the
C/O core white dwarfs in post-common-envelope binaries to be in longer period systems.

The period distribution of the double white dwarfs presented in Figure
\ref{fig:mp} shows a trend with mass, at least in the observed lower limit in period.
The shortest period binaries with $\sim0.4 M_{\odot}$ white dwarfs are in 0.1 d systems,
whereas the shortest period 0.2-0.3 $M_{\odot}$ white dwarfs are in 0.01 d systems.
The median period decreases from 0.64 d to 0.24 d for
$M=0.3-0.5 M_{\odot}$ to $M<0.3 M_{\odot}$ white dwarfs in Figure \ref{fig:mp}.
With masses ranging from 0.29 to 0.43 $M_{\odot}$, the period distribution for
the four binaries presented in this paper is consistent with the
period distribution of the post-common-envelope binaries presented here. The rest of
the low-mass white dwarfs in our sample are also likely in binary systems with $\sim$day
long periods. However, our Gemini snapshot survey is not sensitive to such long periods.

\citet{zorotovic11} looked for a significant correlation between the white dwarf mass and
orbital period for He-core and C/O-core white dwarfs separately, but they could not reject
the null hypothesis based on an F-test. Adding the low-mass white dwarfs from this paper, the ELM Survey,
and the literature (the sample shown in Figure \ref{fig:mp}) does not change the results;
we still cannot reject the null hypothesis (no correlation with mass) based on an F-test
of the current binary white dwarf sample. This is almost certainly due to the fact that
the sample of long period systems with high masses is incomplete, as it
is relatively hard to identify these systems and constrain their parameters.
The population synthesis calculations \citep{nelemans05} predict many low-mass ($\sim0.4 M_{\odot}$) white dwarfs
at $P\sim10$ d, yet they are missing from the observational samples. Hence, larger
samples of binary white dwarfs that include longer period systems are needed to definitively
find a trend between orbital period and primary white dwarf mass.

All four binary systems presented in this paper will merge within a Hubble
time, with total masses of $\geq 0.76, 0.52, 0.68,$ and $0.68 M_{\odot}$, respectively.
The quickest merger system is J1235+1543, which contains a 0.35 $M_{\odot}$ white dwarf
with a $M\geq0.17 M_{\odot}$ companion. Note that a 0.17 $M_{\odot}$
companion is expected to form after the 0.35 $M_{\odot}$ white dwarf and it should be
brighter. Hence, this is likely a low inclination system ($i\leq36^{\circ}$) with a
companion that is comparable to or more massive than 0.35 $M_{\odot}$. 
Based on its distance of 386 pc, J1235+1543 has a gravitational wave strain of
$\log{h} \geq -22.2$ at $\log{\nu} = -3.2$. This strain is comparable to that of the
AM CVn binary GP Com. Hence, J1235+1543 is unlikely to be detected by
{\em LISA} \citep{roelofs07,korol17}.

\section*{Acknowledgements}

This work is in part supported by the NSF and NASA under grants
AST-1312678, AST-1312983, and NNX14AF65G.

\appendix
\section{Radial Velocity Data for Four Binaries}

\begin{table}
\centering
\caption{J0834+3049}
\begin{tabular}{cr}
\hline
HJD$-$2457000 & $v_{helio}$ \\
(days) & (\kms) \\
\hline
389.864816 & 233.3 $\pm$ 40.3 \\
389.869532 & 272.5 $\pm$ 42.5 \\
389.873496 & 316.3 $\pm$ 44.2 \\
389.877461 & 372.8 $\pm$ 48.5 \\
389.881431 & 247.0 $\pm$ 74.7 \\
427.677917 & 46.5 $\pm$ 31.6 \\
427.929727 & 194.4 $\pm$ 27.4 \\
428.675806 & 156.7 $\pm$ 30.0 \\
428.747331 & 370.1 $\pm$ 25.3 \\
428.924687 & -93.3 $\pm$ 45.7 \\
429.694757 & 327.6 $\pm$ 24.1 \\
429.711053 & 227.3 $\pm$ 30.8 \\
429.738483 & 199.7 $\pm$ 27.9 \\
429.902945 & 216.6 $\pm$ 37.5 \\
430.721276 & 38.5 $\pm$ 25.8 \\
430.759435 & 101.7 $\pm$ 29.2 \\
430.836990 & 328.4 $\pm$ 35.2 \\
430.920772 & 236.4 $\pm$ 24.1 \\
\hline
\end{tabular}
\end{table}

\begin{table}
\centering
\caption{J1235+1543}
\begin{tabular}{cr}
\hline
HJD$-$2457000 & $v_{helio}$ \\
(days) & (\kms) \\
\hline
409.083366 & -15.7 $\pm$ 41.2 \\
409.084900 & -82.0 $\pm$ 55.2 \\
409.086433 & -101.0 $\pm$ 33.8 \\
409.088731 & -155.7 $\pm$ 31.6 \\
409.090264 & -190.5 $\pm$ 37.0 \\
409.091798 & -92.3 $\pm$ 26.6 \\
409.093331 & -179.3 $\pm$ 28.6 \\
409.094865 & -203.0 $\pm$ 51.2 \\
409.096398 & -135.2 $\pm$ 28.5 \\
409.097938 & -157.3 $\pm$ 47.3 \\
409.099471 & -39.9 $\pm$ 36.1 \\
427.984125 & -0.8 $\pm$ 23.0 \\
427.987065 & 96.7 $\pm$ 23.3 \\
427.991093 & 156.0 $\pm$ 23.3 \\
427.994080 & 153.2 $\pm$ 23.4 \\
427.997043 & 124.2 $\pm$ 22.8 \\
427.999994 & 114.7 $\pm$ 30.8 \\
428.002992 & 7.5 $\pm$ 24.4 \\
428.005932 & -78.0 $\pm$ 27.3 \\
428.009891 & -132.9 $\pm$ 26.1 \\
428.012842 & -150.7 $\pm$ 25.6 \\
428.015794 & -107.2 $\pm$ 24.4 \\
428.018769 & -45.9 $\pm$ 27.0 \\
429.002616 & -136.6 $\pm$ 23.9 \\
429.005579 & -156.9 $\pm$ 22.2 \\
429.009631 & -71.3 $\pm$ 22.9 \\
429.012594 & -14.6 $\pm$ 23.9 \\
429.983303 & 33.5 $\pm$ 21.9 \\
429.986752 & -24.3 $\pm$ 22.7 \\
430.989533 & -119.1 $\pm$ 22.5 \\
430.992034 & -63.1 $\pm$ 21.7 \\
430.993608 & -61.4 $\pm$ 21.6 \\
430.995194 & -11.2 $\pm$ 21.7 \\
430.996791 & 36.5 $\pm$ 23.1 \\
430.998353 & 109.5 $\pm$ 22.5 \\
430.999916 & 143.2 $\pm$ 22.3 \\
431.001490 & 156.1 $\pm$ 23.3 \\
431.003053 & 188.2 $\pm$ 23.1 \\
431.004615 & 185.5 $\pm$ 24.9 \\
\hline
\end{tabular}
\end{table}

\begin{table}
\centering
\caption{J1237+4913}
\begin{tabular}{cr}
\hline
HJD$-$2457000 & $v_{helio}$ \\
(days) & (\kms) \\
\hline
391.101456 & -176.3 $\pm$ 47.2 \\
391.105073 & -52.3 $\pm$ 43.8 \\
391.109443 & -101.3 $\pm$ 46.5 \\
391.116495 & 14.4 $\pm$ 39.6 \\
391.120112 & 50.2 $\pm$ 65.7 \\
391.123731 & 52.9 $\pm$ 36.4 \\
575.756677 & -41.1 $\pm$ 39.1 \\
575.763138 & -100.6 $\pm$ 38.8 \\
575.768839 & -165.5 $\pm$ 41.0 \\
575.774539 & -116.7 $\pm$ 33.9 \\
575.780997 & -179.7 $\pm$ 37.2 \\
575.786688 & -153.4 $\pm$ 35.7 \\
575.792398 & -160.1 $\pm$ 34.2 \\
575.798854 & -154.0 $\pm$ 35.4 \\
575.804554 & -132.3 $\pm$ 34.4 \\
575.810255 & -155.8 $\pm$ 42.1 \\
575.815820 & -43.7 $\pm$ 45.5 \\
580.760275 & -72.1 $\pm$ 42.7 \\
580.766731 & -103.7 $\pm$ 38.5 \\
580.772431 & 64.7 $\pm$ 41.8 \\
580.778131 & 93.6 $\pm$ 40.7 \\
580.784589 & 125.5 $\pm$ 41.0 \\
580.790289 & 91.0 $\pm$ 42.5 \\
580.795990 & 124.9 $\pm$ 35.9 \\
580.802447 & 65.0 $\pm$ 41.5 \\
580.808148 & 39.6 $\pm$ 44.1 \\
747.010077 & -55.8 $\pm$ 35.0 \\
747.019614 & -207.8 $\pm$ 36.7 \\
747.028921 & -249.9 $\pm$ 43.9 \\
814.913842 & -24.6 $\pm$ 44.1 \\
814.981547 & -3.9 $\pm$ 48.1 \\
815.009915 & 147.8 $\pm$ 54.1 \\
838.830221 & -194.2 $\pm$ 37.8 \\
838.860139 & -138.1 $\pm$ 35.5 \\
838.882893 & 8.4 $\pm$ 34.5 \\
838.915161 & -28.2 $\pm$ 33.8 \\
838.940079 & -131.3 $\pm$ 35.4 \\
839.620038 & -97.1 $\pm$ 36.7 \\
839.674574 & 23.5 $\pm$ 34.1 \\
839.868874 & 105.1 $\pm$ 33.6 \\
840.861070 & -69.6 $\pm$ 39.9 \\
\hline
\end{tabular}
\end{table}

\begin{table}
\centering
\caption{J2342+0811}
\begin{tabular}{cr}
\hline
HJD$-$2457000 & $v_{helio}$ \\
(days) & (\kms) \\
\hline
582.052454 & -136.3 $\pm$ 28.3 \\
582.058911 & -168.4 $\pm$ 37.0 \\
583.040117 & 75.2 $\pm$ 30.7 \\
583.046574 & 97.2 $\pm$ 28.3 \\
583.052275 & 86.0 $\pm$ 37.6 \\
583.057977 & -38.4 $\pm$ 28.9 \\
583.064433 & -50.1 $\pm$ 38.8 \\
691.682281 & -31.5 $\pm$ 24.6 \\
691.687605 & -13.9 $\pm$ 30.2 \\
691.733320 & -118.8 $\pm$ 28.8 \\
691.738377 & -39.1 $\pm$ 34.1 \\
691.744743 & -85.6 $\pm$ 28.3 \\
691.750495 & -34.1 $\pm$ 27.9 \\
691.757126 & -14.4 $\pm$ 25.7 \\
691.765158 & 32.6 $\pm$ 24.4 \\
691.772716 & 42.8 $\pm$ 27.2 \\
691.781859 & 70.9 $\pm$ 24.5 \\
691.790238 & 147.4 $\pm$ 26.2 \\
691.799589 & 126.0 $\pm$ 24.5 \\
723.575083 & 31.7 $\pm$ 24.0 \\
723.616491 & -170.0 $\pm$ 32.7 \\
723.664692 & 22.0 $\pm$ 24.5 \\
723.720335 & 95.2 $\pm$ 26.8 \\
723.764336 & 4.0 $\pm$ 58.4 \\
724.567698 & 154.2 $\pm$ 28.4 \\
\hline
\end{tabular}
\end{table}

\end{document}